\documentclass[11pt]{article}
\usepackage{graphicx}
\usepackage{latexsym,graphicx}
\usepackage{float}
\usepackage{amsmath}
\usepackage{amsfonts}
\usepackage{amssymb}
\usepackage{epstopdf}
\usepackage{color}
\DeclareGraphicsExtensions{.eps}
\usepackage[left=2.5cm,top=2.5cm,right=2.5cm,bottom=2.5cm]{geometry}

\catcode `\@=11
\catcode `\@=12
\begin{document}
	\begin{center}
		\large{\bf{Dark Energy Nature in Logarithmic $f(R,T)$ Cosmology}} \\
		\vspace{5mm}
		\normalsize{ Dinesh Chandra Maurya$^{1}$,  Jagat Singh$^{2}$, Lalit Kumar Gaur$^{3}$}\\
		\vspace{5mm}
		\normalsize{$^{1}$ Centre for Cosmology, Astrophysics and Space Science, GLA University, Mathura-281 406,
		Uttar Pradesh, India.}\\
		\vspace{5mm}
		\normalsize{$^{2}$ G. B. Pant DSEU Okhla Campus-III, Delhi, Sector-9, Dwarka, New Delhi-110077, India.}\\
		\vspace{5mm}
		\normalsize{$^{3}$Department of Physics, Seth Gyaniram Bansidhar Podar College, Nawalgarh-333042 (Jhunjhunu), Rajsthan, India.}\\
			\vspace{2mm}
		{$^{1}$Email:dcmaurya563@gmail.com}\\
			\vspace{2mm}
		{$^{2}$Email:jagatiitdelhi@gmail.com}\\
			\vspace{2mm}
		{$^{3}$E-mail:gaurlalit520@gmail.com}\\
	\end{center}
	\vspace{5mm}
	%\date{}
	%\maketitle
	%%%%%%%%%%%%%%%%%%%%%%%%%%%%%%%%%%%%%%%%%%%%%%%%%%%%%%%%%%%%%%%%%%%%%%%%%%%%%%%%%%%%%%%%%%%%
	\begin{abstract}
The present research paper is an investigation of dark energy nature of logarithmic $f(R, T)$-gravity cosmology in a flat FLRW space-time universe. We have derived modified Einstein's field equations for the function $f(R, T)=R-16\pi G\alpha\ln(T)$ where $R$ is the Ricci scalar curvature, $T$ is the trace of the stress energy momentum tensor and $\alpha$ is a model parameter. We have solved field equations in the form of two fluid scenario as perfect-fluid and dark-fluid, where dark fluid term is derived in the form of perfect fluid source. We have made an observational constraints on the cosmological parameters $\Omega_{(m)}, \omega^{(de)}$ and $H_{0}$ using $\chi^{2}$ test with observational datasets like Pantheon sample of SNe Ia and $H(z)$. With these constraints we have discussed our model with deceleration parameter $q$, energy parameters $\Omega_{(m)}, \Omega_{(de)}$, EoS parameter $\omega^{(de)}$ etc. Also, we have done Om diagnostic analysis. The derived $f(R, T)$ model shows a quintessence dark energy model $\omega^{(de)}>-1$ and late-time universe approaches to $\Lambda$CDM model.
\end{abstract}
	\smallskip
	\vspace{5mm}
	%\date{}
	%\maketitle
	%\vspace{1cm}

	%%%%%%%%%%%%%%%%%%%%%%%%%%%%%%%%%%%%%%%%
	\section{Introduction}
The noble discovery in \cite{ref1}-\cite{ref15} approves the cosmic acceleration in expansion of the universe. The classical General Relativity (GR) predicts the expansion of the universe and it suggests that the expansion should be decelerating with time. But the observations in \cite{ref1}-\cite{ref15} suggest that the current universe has entered in a second phase of accelerated expansion which is started around redshift $z=1$. Also, it is observed that approximately $70\%$ of the total energy density of the universe is in some mysterious form called ``Dark Energy" which has high negative pressure that creates repulsive forces among the galaxies and results the accelerating expansion of the universe. But nobody knows actual nature of the Dark Energy. Einstein obtained this acceleration in his cosmological model by adding a constant term $\Lambda$, called ``Cosmological Constant". Although the ``Cosmological Constant $\Lambda-$term" is the best fit candidate for dark energy, but it has two problems, first is about its origin and second is fine-tuning its value with dark energy. To solve the dark energy problem and cosmological constant problem, in literature several modified and alternative theories of gravity to GR are presented by the cosmologists time to time but the dark energy problem is an unsolved problem till to date.\\

Current studies focus on the determination of the equation of state parameter $\omega$ (see the references \cite{ref16, ref17, ref18, ref19}) to measure the properties of dark energy component of the universe from observational data. The equation of state parameter $\omega$ is defined as the ratio of pressure to the energy density of the fluid $\omega(t)=\frac{p}{\rho}$ and is not necessarily constant. The vacuum energy having EoS $\omega=-1$ is the simplest dark energy candidate and is equivalent to ``Cosmological Constant $\Lambda$-term". Alternatives to vacuum energy can be described by minimally coupled scalar fields, are quintessence $(\omega > -1)$, phantom energy $(\omega <-1)$ and Quintom (that can across from phantom region to quintessence region as evolved) and have time dependent EoS parameter. Some observational constraints on limits of EoS $\omega$ are obtained by Knop et al. \cite{ref20} and Tegmark et al. \cite{ref21} as $-1.67<\omega<-0.62$ and $-1.33<\omega<-0.79$ respectively. The latest results on limit of EoS are obtained as $-1.44<\omega<-0.92$ at $68\%$ confidence level in 2009 by Hinshaw et al. \cite{ref22}; Komatsu et al. \cite{ref23}. However, we are not on a stage to use a constant value of $\omega$ because we have not observational evidences which makes a distinction between constant and variable $\omega$. A large number of cosmologists, considered the equation of state parameter as a constant (Kujat et al. \cite{ref24}; Bartelmann et al. \cite{ref25}) with phase wise value $-1, 0, +\frac{1}{3}$ and $+1$ for vacuum fluid, dust fluid, radiation and stiff dominated universe, respectively. But generally, $\omega$ is time or redshift dependent function (Jimenez \cite{ref27}; Das et al. \cite{ref28}). In literature, several cosmologists (\cite{ref29}-\cite{ref37}) have presented cosmological models with variable EoS parameter $\omega$.\\

Various $f(R)$ theory applications to cosmology and gravity, including inflation, dark energy, local gravity constraints, cosmological perturbations, and spherically symmetric solutions in weak and strong gravitational backgrounds, are reviewed in \cite{ref38}. In \cite{ref39,ref40,ref41,ref42}, a review of several well-established topics and the most recent advances in modified gravity in cosmology are presented, with an emphasis on inflation, bouncing cosmology, and the late-time acceleration era employing $F(R)$, $F(G)$, and $F(T)$ gravity theories. In the context of higher order theories of gravity, the issues of quintessence and cosmic acceleration have been covered in \cite{ref43}. In \cite{ref44}, a review of dynamical dark energy models is offered. In the context of modified gravity theories with negative and positive curvatures, references \cite{ref45} and \cite{ref46} seek to unify inflation with cosmic acceleration. In \cite{ref47}, a variety of workable $F(R)$ gravity dark energy theories are examined.\\

A generalization of $f(R)$ gravity by including the trace $T$ of stress-energy-momentum tensor $T_{ij}$ has been proposed by Harko {\it et al.} \cite{ref48} known as $f(R, T)$ gravity. The different cosmological and astrophysical aspects of $f(R, T)$ gravity have been extensively studied by several authors. Several authors \cite{ref49} have investigated the physical and geometrical aspects of modified $f(R, T)$ cosmological models in different context. The accelerated expansion phase of the universe plays an important role in the dynamical history of our universe. Using different forms of the $f(R, T)$ gravity, Harko {\it et al.} \cite{ref48} have constructed some FLRW modified cosmological models. Some generalization of $F(R)$ and $F(T)$ gravity theories are studied by Myrzakulov \cite{ref50} and on the basis of this Lagrangian, he derived the field equations in $f(R, T)$ gravity and have obtained some exact solutions for the specific $F(R,T) = \mu R + \nu T$ function. After that several cosmological models are proposed in $f(R, T)$ gravity \cite{ref51}-\cite{ref81}.\\

The first logarithmic $f(R, T)$ gravity theory has been proposed by Elizalde et al.\cite{ref82} in the form of $f(R, T)= R+ \alpha R^2+2\beta \ln(T)$ in which they have studied the energy and stability conditions of the cosmological model. Recently Deb and Deshamukhya \cite{ref83} have studied some constraints on simple form of logarithmic $f(R, T)= R+ 16\pi G \alpha \ln (T)$ gravity by using dark energy parameters and Hubble constant $H_{0}$. Here, we have studied the behaviour of dark energy parameters and equation of state parameters in logarithmic $f(R, T)= R-16\pi G \alpha \ln (T)$ gravity with observational constraints. The motivation behind choosing such a type of specific $f(R,T)$ function is that in this case, all energy conditions are satisfied. The testing of energy conditions of such type cosmological models are studied in details in \cite{ref84}.\\

The present paper is organized as follows: Sect.~1 is introductory, Sect.~2 contains formulation of modified field equations for $f(R,T)= R-16\pi G\alpha \ln(T)$ and its solution. In Sect.~3, we have made observational constraints on energy parameters, Sect.~4 contains discussion of results with Om diagnostic analysis. In last section 5 have concluding remarks.
%%%%%%%%%%%%%%%%%%%%%%%%%%%%%%%%%%%%%%%%%%%%%%%%%%%%%%%%%%%%%%%%%%%%%%%%%%%%%%%
%%%%%%%%%%%%%%%%%%%%%%%%%%%%%%%%%%%%%%%%%%%%%%%%%%%%%%%SECTION 2
%%%%%%%%%%%%%%%%%%%%%%%%%%%%%%%%%%%%%%%%%%%%%%%%%%%%%%%%%%%%%%%%%%%%%%%%%%%%%%%
\section{Field Equations for Logarithmic $f(R, T)$-Gravity and Solution}

We consider the action for the logarithmic $f(R,T)= R-16\pi G\alpha\ln(T)$ function as,

\begin{equation}\label{eq1}
    S = \int\sqrt{-g}\left[\frac{R}{16 \pi G}-\alpha\ln(T)+L_m\right]d^4x,
\end{equation}
where $L_m$ is the matter Lagrangian, $R$ is the Ricci scalar curvature, $T$ is the trace of the matter stress-energy momentum tensor $T_{ij}$ and $\alpha$ is the model parameter. The motivation behind choosing such a type of specific $f(R,T)$ function is that in this case, all energy conditions are satisfied. The testing of energy conditions of such type cosmological models are studied in details in \cite{ref84}.\\

Variation of action (\ref{eq1}) with respect to metric tensor $g_{ij}$, we obtain the following field equations,
\begin{equation}\label{eq2}
  R_{ij} - \frac{1}{2}g_{ij}R = 8 \pi G\left[T_{ij}+ T_{ij}^{(de)}\right],
\end{equation}
where
\begin{equation}\label{eq3}
    T_{ij}^{(de)}=-\frac{2\alpha}{T}\left( T_{ij} + \frac{T}{2} g _{ij}\ln T + \Theta_{ij}\right),
\end{equation}
where the term $\Theta_{ij}$, which plays a crucial role in $f(R,T)$ gravity as it contains matter Lagrangian $L_m$, is given by
\begin{equation}\label{eq4}
    \Theta_{ij}= g^{\beta\gamma} \frac{\delta T_{\beta\gamma}}{\delta g^{ij}} = -2 T_{ij} + g_{ij}L_m - 2 \frac{\delta^2 L_m}{\delta g^{ij} \delta. g^{\beta\gamma}}
\end{equation}
Clearly, depending on the nature of the matter field, the field equation for $f(R,T)$ gravity will be different. Now, assuming the Universe is filled with perfect fluid, the stress-energy-momentum tensor is
\begin{equation}\label{eq5}
    T_{ij}=(\rho+p)u_{i}u_{j}-pg_{ij},
\end{equation}
where $\rho$ is the energy density, $p$ is the isotropic pressure of the perfect fluid source and $u^{i}=(1, 0, 0, 0)$ is four fundamental velocity in co-moving coordinates and the matter Lagrangian density can be assumed as $L_m=-p$. Now, we consider the Friedmann-Lemaitre-Robertson-Walkar (FLRW) metric in spherical coordinate for flat Universe as,
\begin{equation}\label{eq6}
     ds^2= c^{2}dt^2 - a(t)^2 [dx^2+dy^{2}+dz^{2}],
\end{equation}
where $a(t)$ denotes scale factor of the Universe.

Now, assuming $8\pi G=1~~\&~~c=1$ in cosmological units, we get the field equations for the metric (\ref{eq6}) as,
\begin{equation}\label{eq7}
   3H^2 = \rho+\rho^{(de)}
\end{equation}
and
\begin{equation}\label{eq8}
  2\dot{H}+3H^{2}=-p-p^{(de)}
\end{equation}
where
\begin{equation}\label{eq9}
  \rho^{(de)}=\frac{2\alpha(\rho+p)}{T}-\alpha\ln(T),~~~~p^{(de)}=\alpha\ln(T)
\end{equation}
respectively called as dark energy density and corresponding isotropic pressure. Here $H$ is the Hubble parameter defined by $H=\frac{\dot{a}}{a}$, and the trace $T$ of stress-energy momentum tensor is given as $T=\rho-3p$.

The equation of continuity is obtained as
\begin{equation}\label{eq10}
  \dot{\rho}+3H(\rho+p)+[\dot{\rho}^{(de)}+3H(\rho^{(de)}+p^{(de)})]=0
\end{equation}
Taking non-interacting condition
\begin{equation}\label{eq11}
  \dot{\rho}+3H(\rho+p)=0,~~~~~~\dot{\rho}^{(de)}+3H(\rho^{(de)}+p^{(de)})=0
\end{equation}
Now, taking the equation of state (EoS) as $p=\omega\rho$ with $\omega=$constant, integrating Eq.~(\ref{eq11}), we get
\begin{equation}\label{eq12}
  \rho=\rho_{0}\left(\frac{a_{0}}{a}\right)^{3(1+\omega)},~~~~~~\rho^{(de)}=\rho_{0}^{(de)}\left(\frac{a_{0}}{a}\right)^{3(1+\omega^{(de)})}
\end{equation}
Now, from equation (\ref{eq7}), we obtain
\begin{equation}\label{eq13}
  \Omega_{(m)}+\Omega_{(de)}=1
\end{equation}
where $\Omega_{(m)}=\frac{\rho}{3H^{2}}$ and $\Omega_{(de)}=\frac{\rho^{(de)}}{3H^{2}}$ are respectively known as matter energy density parameter and dark energy density parameter.

From Eqs.~(\ref{eq12}) \& (\ref{eq13}), we get the Hubble function as
\begin{equation}\label{eq14}
  H=H_{0}\sqrt{\Omega_{(m)0}\left(\frac{a_{0}}{a}\right)^{3(1+\omega)}+\Omega_{(de)0}\left(\frac{a_{0}}{a}\right)^{3(1+\omega^{(de)})}}
\end{equation}
or
\begin{equation}\label{eq15}
  H=H_{0}\sqrt{\Omega_{(m)0}(1+z)^{3(1+\omega)}+\Omega_{(de)0}(1+z)^{3(1+\omega^{(de)})}}
\end{equation}
From Eqs.~(\ref{eq7}), (\ref{eq8}) \& (\ref{eq9}), we get the expression for deceleration parameter as
\begin{equation}\label{eq16}
  q=\frac{1}{2}+\frac{3}{2}\frac{p+\alpha\ln(T)}{\rho+\frac{2\alpha(\rho+p)}{T}-\alpha\ln(T)}
\end{equation}
where
\begin{equation}\label{eq17}
  \alpha=\frac{\rho_{0}^{(de)}}{\frac{2(1+\omega)}{1-3\omega}-\ln(1-3\omega)-\ln(\rho_{0})}
\end{equation}
%%%%%%%%%%%%%%%%%%%%%%%%%%%%%%%%%%%%%%%%%%%%%%%%%%%%%%%%%%%%%%%%%%%%%%%%%%%%%%%%%%%%%%%%%%%%%%%%%
%%%%%%%%%%%%%%%%%%%%%%%%%%%%%%%%%%%%%%%%%%%%%%%%%%%%%%%SECTION 3
%%%%%%%%%%%%%%%%%%%%%%%%%%%%%%%%%%%%%%%%%%%%%%%%%%%%%%%%%%%%%%%%%%%%%%%%%%%%%%%%%%%%%%%%%%%%%%%%%
\section{Observational Constraints}
Current theoretical cosmology is focused on best-fitting of the cosmological parameters with observational cosmology. Hence, we have obtained the best curve of Hubble parameter $H(z)$ and apparent magnitude $m(z)$ using observational datasets $H(z)$, union 2.1 compilation and Pantheon datasets of SNe Ia observations by applying $\chi^{2}$-test given as follows:
\begin{equation}\label{eq18}
  \chi^{2}=\sum_{i=1}^{i=N}\frac{[O_{i}-E_{i}]^{2}}{\sigma_{i}^{2}}
\end{equation}
where $N$ denotes the number of data, $O_{i},~E_{i}$ represent the observed and estimated datasets respectively and $\sigma_{i}$ denotes standard deviations.
%%%%%%%%%%%%%%%%%%%%%%%%%%%%%%%%%%%%%%%%%%%%%%%%%%%%%%%%%%%%%%%%%%%%%%%%%%%%%%%%
\subsection*{Hubble Parameter}
%%%%%%%%%%%%%%%%%%%%%%%%%%%%%%%%%%%%%%%%%%%%%%%%%%%%%%%%%%%%%%%%%%%%%%%%%
\begin{table}[H]
  \centering
  {\tiny
  \begin{tabular}{|c|c|c|c|c|c|c|c|c|c|c|c|}
     \hline
      % after \\: \hline or \cline{col1-col2} \cline{col3-col4} ...
			S.No. & $z$   & $H(z)$  & $\sigma_{H}$  & Reference & Method  & S.No. & $z$   & $H(z)$  & $\sigma_{H}$  & Reference & Method\\
			\hline
			1  & $0$      & $67.77$ & $1.30$   & \cite{ref85}   & DA      & 24  & $0.4783$  & $80.9$  & $9$     & \cite{ref99}  & DA\\
			2  & $0.07$   & $69$    & $19.6$   & \cite{ref86}   & DA      & 25  & $0.48$    & $97$    & $60$    & \cite{ref87}  & DA\\
			3  & $0.09$   & $69$    & $12$     & \cite{ref98}   & DA      & 26  & $0.51$    & $90.4$  & $1.9$   & \cite{ref89}  & DA\\
			4  & $0.10$   & $69$    & $12$     & \cite{ref87}   & DA      & 27  & $0.57$    & $96.8$  & $3.4$   & \cite{ref100}  & DA\\
			5  & $0.12$   & $68.6$  & $26.2$   & \cite{ref86}   & DA      & 28  & $0.593$   & $104$   & $13$    & \cite{ref97}  & DA\\
			6  & $0.17$   & $83$    & $8$      & \cite{ref87}   & DA      & 29  & $0.60$    & $87.9$  & $6.1$   & \cite{ref91}  & DA\\
			7  & $0.179$  & $75$    & $4$      & \cite{ref97}   & DA      & 30  & $0.61$    & $97.3$  & $2.1$   & \cite{ref89}  & DA\\
			8  & $0.1993$ & $75$    & $5$      & \cite{ref97}   & DA      & 31  & $0.68$    & $92$    & $8$     & \cite{ref97}  & DA\\
			9  & $0.2$    & $72.9$  & $29.6$   & \cite{ref86}   & DA      & 32  & $0.73$    & $97.3$  & $7$     & \cite{ref91}  & DA\\
		   10  & $0.24$   & $79.7$  & $2.7$    & \cite{ref88}   & DA      & 33  & $0.781$   & $105$   & $12$    & \cite{ref97}  & DA\\
		   11  & $0.27$   & $77$    & $14$     & \cite{ref87}   & DA      & 34  & $0.875$   & $125$   & $17$    & \cite{ref97}  & DA\\
		   12  & $0.28$   & $88.8$  & $36.6$   & \cite{ref86}   & DA      & 35  & $0.88$    & $90$    & $40$    & \cite{ref87}  & DA\\
		   13  & $0.35$   & $82.7$  & $8.4$    & \cite{ref90}   & DA      & 36  & $0.9$     & $117$   & $23$    & \cite{ref87}  & DA\\
		   14  & $0.352$  & $83$    & $14$     & \cite{ref97}   & DA      & 37  & $1.037$   & $154$   & $20$    & \cite{ref88}  & DA\\
		   15  & $0.38$   & $81.5$  & $1.9$    & \cite{ref89}   & DA      & 38  & $1.3$     & $168$   & $17$    & \cite{ref87}  & DA\\
		   16  & $0.3802$ & $83$    & $13.5$   & \cite{ref90}   & DA      & 39  & $1.363$   & $160$   & $33.6$  & \cite{ref93}  & DA\\
           17  & $0.4$    & $95$    & $17$     & \cite{ref98}   & DA      & 40  & $1.43$    & $177$   & $18$    & \cite{ref87}  & DA\\
           18  & $0.004$  & $77$    & $10.2$   & \cite{ref99}   & DA      & 41  & $1.53$    & $140$   & $14$    & \cite{ref87}  & DA\\
           19  & $0.4247$ & $87.1$  & $11.2$   & \cite{ref99}   & DA      & 42  & $1.75$    & $202$   & $40$    & \cite{ref93}  & DA\\
           20  & $0.43$   & $86.5$  & $3.7$    & \cite{ref88}   & DA      & 43  & $1.965$   & $186.5$ & $50.4$  & \cite{ref88}  & DA\\
	       21  & $0.44$   & $82.6$  & $7.8$    & \cite{ref91}   & DA      & 44  & $2.3$     & $224$   & $8$     & \cite{ref96}  & DA\\
           22  & $0.44497$& $92.8$  & $12.9$   & \cite{ref99}   & DA      & 45  & $2.34$    & $222$   & $7$     & \cite{ref94}  & DA\\
           23  & $0.47$   & $89$    & $49.6$   & \cite{ref92}   & DA      & 46  & $2.36$    & $226$   & $8$     & \cite{ref95}  & DA\\
     \hline
   \end{tabular}
    \caption{Hubble's constant table.}\label{T1}}
\end{table}
%%%%%%%%%%%%%%%%%%%%%%%%%%%%%%%%%%%%%%%%%%%%%%%
The Hubble parameter $H$ is one of the important observational cosmological parameter which reveals the rate of expansion of the universe. We have considered $46$ $H(z)$ datasets with redshift $z$ (see the Table 1) estimated using Differential Age (DA) method by cosmologists time to time in \cite{ref85}-\cite{ref100} for best curve-fitting of $H(z)$. Here, we have considered matter dominated universe with $\omega=0$, hence, the Eq.~(\ref{eq15}) becomes
\begin{equation}\label{eq19}
  H(z)=H_{0}\sqrt{\Omega_{(m)0}(1+z)^{3}+\Omega_{(de)0}(1+z)^{3(1+\omega^{(de)})}}
\end{equation}
The best fit values of energy parameters are mentioned in Table $2$ and the best fit curve is given by figure 1.\\
%%%%%%%%%%%%%%%%%%%%%%%%%%%%%%%%%%%%%%%%%%%%%%%%%%%%%%%%%%%%%%%%%%%%%%
\begin{figure}[H]
\centering
	\includegraphics[width=10cm,height=8cm,angle=0]{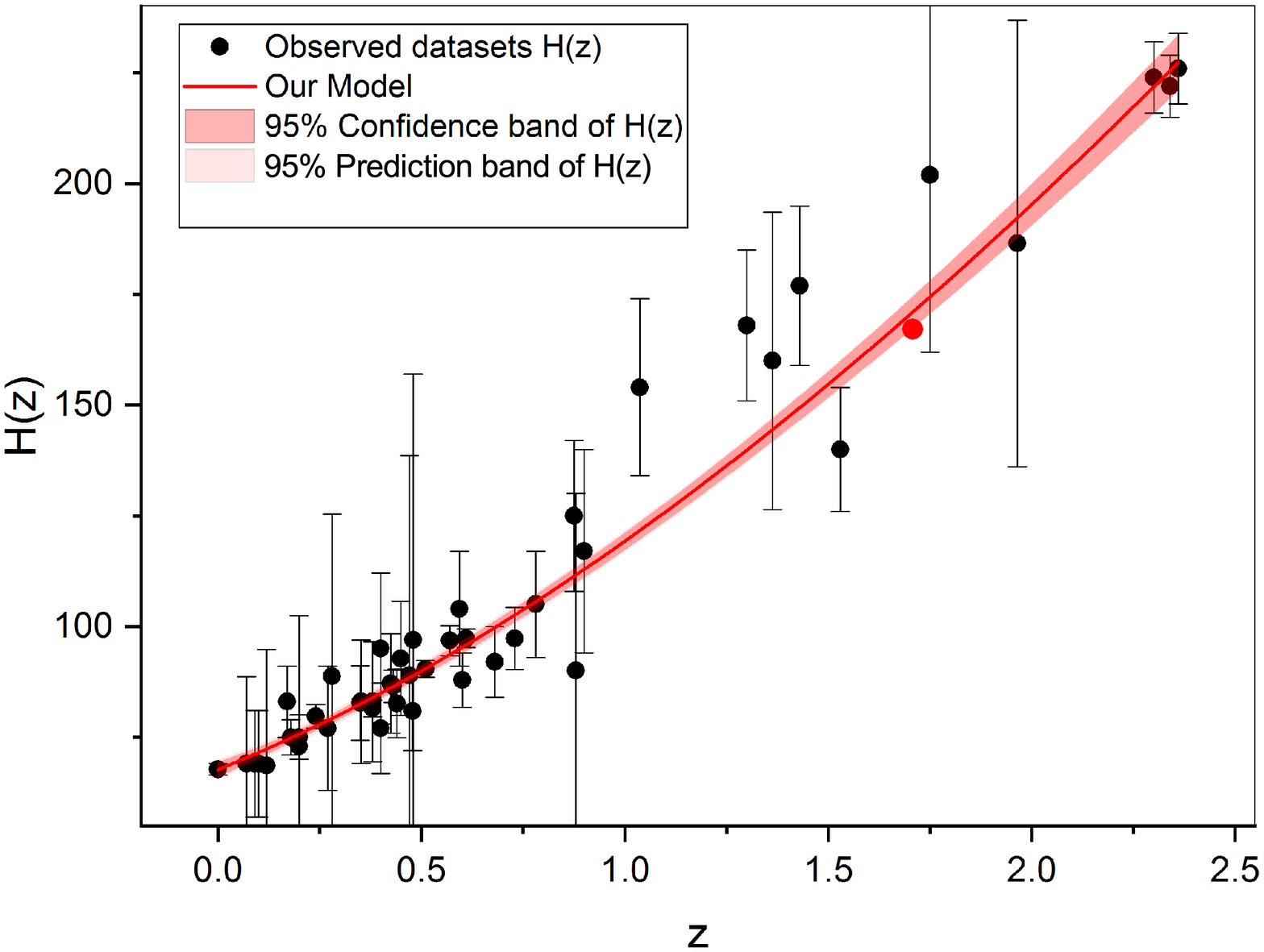}
  	\caption{The best fit curve of Hubble parameter $H(z)$.}
\end{figure}
%%%%%%%%%%%%%%%%%%%%%%%%%%%%%%%%%%%%%%%%%%%%%%%%%%%%%%%%%%%%%%%%%%%%%%

%%%%%%%%%%%%%%%%%%%%%%%%%%%%%%%%%%%%%%%%%%%%%%%%%%%%%%%%%%%%%%%%%%%%%%%%%%%%%%%%%%%%%%
\subsection*{Apparent Magnitude}
We have considered $40$ SNe Ia bined data of $m(z)$ from compilation of supernovae pantheon samples in the range of ($ 0 \leq  z \leq 1.7$ ) \cite{ref101,ref102}. We use the $\chi^{2}$ test formula to achieve the best fit curve for theoretical and empirical results. The expression for apparent magnitude is taken by
\begin{equation}\label{eq20}
  m(z)=16.08+5\times\log_{10}\left(\frac{H_{0}D_{L}}{0.026 c\text{Mpc}}\right).
\end{equation}
where the Luminosity distance $D_{L}$ is given by
\begin{equation}\label{eq21}
D_{L}=c(1+z)\int_{0}^{z}\frac{dz}{H(z)}
\end{equation}
where $c$ is the velocity of light and $H(z)$ is the Hubble parameter given in Eq. (\ref{eq19}).\\

The best fit values of the energy parameters are given in Table 2 and the best fit curve is represented by the figure 2.

%%%%%%%%%%%%%%%%%%%%%%%%%%%%%%%%%%%%%%%%%%%%%%%%%%%%%%%%%%%%%%%%%%%
\begin{table}[H]
  \centering
  \begin{tabular}{|c|c|c|}
     \hline
     % after \\: \hline or \cline{col1-col2} \cline{col3-col4} ...
     Parameters & Bined data & $H(z)$ data \\
     \hline
        $\Omega_{(m)}$    &  $0.2981\pm0.08921$ & $0.26535\pm0.01254$  \\
        $\omega^{(de)}$   &  $-0.81\pm0.22149$   & $-0.58155\pm0.16941$ \\
         $H_{0}$           &  $-$              & $67.67851\pm0.86949$ Km/s/Mpc \\
        $\chi^{2}$        &  $0.01702$        & $0.51113$ \\
        $R^2$             &  $0.99785$        & $0.98385$ \\
     \hline
   \end{tabular}
  \caption{The best-fit values of energy parameters along two data sets SNe Ia and Hubble Parameter $H(z)$.}\label{T2}
\end{table}
%%%%%%%%%%%%%%%%%%%%%%%%%%%%%%%%%%%%%%%%%%%%%%%%%%%%%%%%%%%%%%%%%%%%%%%%
\begin{figure}[H]
\centering
	\includegraphics[width=10cm,height=8cm,angle=0]{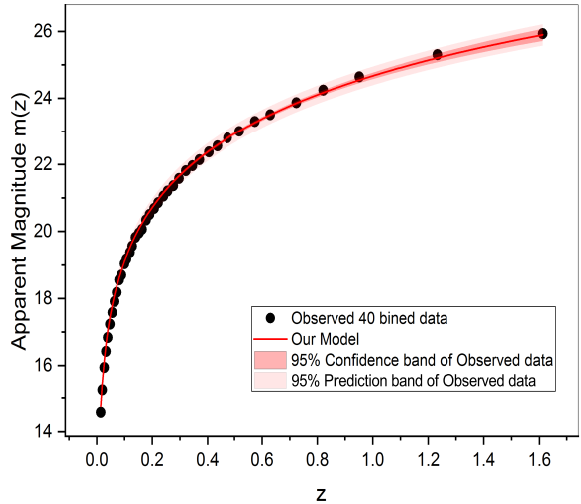}
  	\caption{The best fit curve of apparent magnitude $m(z)$.}
\end{figure}
%%%%%%%%%%%%%%%%%%%%%%%%%%%%%%%%%%%%%%%%%%%%%%%%%%%%%%%%%%%%%%%%%%%%%%%%%%%%%%%%%%%%
The value of Hubble constant measuring from velocity and distances of galaxies is reported as $73\pm1$ Km/s/Mpc that is called late-time version and from another way using ``early time" information, astrophysicists predict that the Hubble constant should be about $67.5 \pm 0.5$ Km/s/Mpc. The values obtained from these two ways are not consistent and this problem is called Hubble tension. The estimated present value of Hubble constant in the derived model is $67.67851\pm0.86949$ Km/s/Mpc which is very closed to ``early time" version of the Hubble constant value and hence, this model may resolve the Hubble tension issue in cosmology, since it is consistent with both early and late-time universe. Niedermann and Sloth \cite{ref103} are reported Hubble constant value $69.6^{+1.0}_{-1.3}$Km/s/Mpc (at $68\%$ C.L.) without the local measurement of the Hubble parameter, bringing the tension down to $2.5~\sigma$.
%%%%%%%%%%%%%%%%%%%%%%%%%%%%%%%%%%%%%%%%%%%%%%%%%%%%%%%%%%%%%%%%%%%%%%%%%%%%%%%%%%%%%%%%%%%%%%%%%%%%%
%%%%%%%%%%%%%%%%%%%%%%%%%%%%%%%%%%%%%%%%%%%%%%%%%%%%%%%%%%%%%%%%%%%%%%%%%%%SECTION 4
%%%%%%%%%%%%%%%%%%%%%%%%%%%%%%%%%%%%%%%%%%%%%%%%%%%%%%%%%%%%%%%%%%%%%%%%%%%%%%%%%%%%%%%%%%%%%%%%%%%%%
\section{Discussion of Results}
\subsection*{Deceleration Parameter}
The expression for deceleration parameter $q$ is given by equation (\ref{eq16}) and its geometrical behaviour is represented by figure 3. One can see that the $q(z)$ is an increasing function of redshift $z$ with signature flipping and it shows a transit phase universe (decelerating to accelerating phase) model. The transition redshift is obtained as $z_{t}=0.6455$ for Pantheon data and $z_{t}=0.7356$ for $H(z)$ data. That is the matter dominated ($\omega=0$) universe is in decelerating phase for $z>z_{t}$ and accelerating for $z<z_{t}$. In literature, Davis et al. \cite{ref104} have obtained the transition redshift $z_{t} \sim 0.6 (1\; \sigma)$ in better agreement with the flat $\Lambda$CDM model ($z_{t} = (2\Omega_{\Lambda}/\Omega_{m})^{\frac{1}{3}} - 1 \sim 0.66$) which is supported our model. The present value of the deceleration parameter is obtained $q_{0}=-0.5276$ for Pantheon data and $q_{0}=-0.5756$ for $H(z)$ data  (see Table 3) which shows that present universe is accelerating phase and is in good agreement with recent observations \cite{ref1}-\cite{ref15}.
%%%%%%%%%%%%%%%%%%%%%%%%%%%%%%%%%%%%%%%%%%%%%%%%%%%%%%%%%%%%%%%%%%%%%%
\begin{figure}[H]
\centering
	\includegraphics[width=8cm,height=7cm,angle=0]{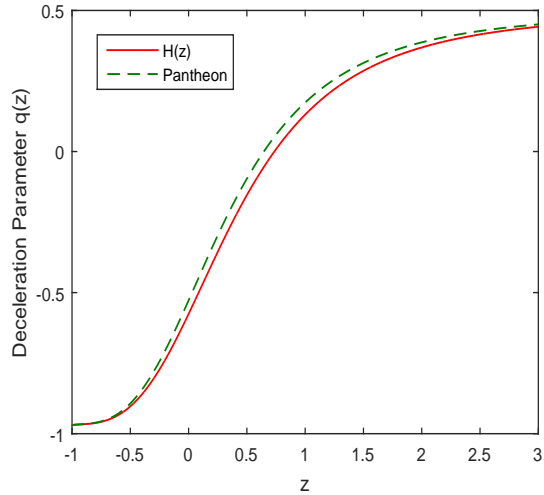}
  	\caption{The behaviour of deceleration parameter $q$ over redshift $z$.}
\end{figure}
%%%%%%%%%%%%%%%%%%%%%%%%%%%%%%%%%%%%%%%%%%%%%%%%%%%%%%%%%%%%%%%%%%%%%%

From equation (\ref{eq8}), we can obtain
\begin{equation}\label{eq22}
  q=\frac{1}{2}+\frac{3}{2}\omega^{(de)}\Omega_{(de)}
\end{equation}
For $q<0$, we have
\begin{equation}\label{eq23}
  \Omega_{(de)}>-\frac{1}{3\omega^{(de)}}
\end{equation}
In our derived model, we have obtained for $q<0$, $\Omega_{(de)}>0.411522634, 0.573180867$ for two datasets and these are in good agreement with observations. Also, for $q=q_{0}$, the energy parameters are $\Omega_{(m)}=0.2981\pm0.08921$, $\Omega_{(de)}=0.7019$ for Pantheon data and $\Omega_{(m)}=0.26535\pm0.01254$, $\Omega_{(de)}=0.73465$ for $H(z)$ datasets.
%%%%%%%%%%%%%%%%%%%%%%%%%%%%%%%%%%%%%%%%%%%%%%%%%%%%%%%%%%%%%%%%%%
\begin{table}[H]
  \centering
  \begin{tabular}{|c|c|c|}
     \hline
     % after \\: \hline or \cline{col1-col2} \cline{col3-col4} ...
     Parameters             & Bined data                             & $H(z)$ data \\
     \hline
        $\Omega_{(de)0}$    &  $0.7019$                              & $0.73465$ \\
        $\rho_{0}$          &  $4.3012\times10^{-36}~ gm/cm^3$       & $3.8286\times10^{-36}~ gm/cm^3$  \\
        $\rho^{(de)}_{0}$   &  $1.0127\times10^{-35}~ gm/cm^3$       & $1.0600\times10^{-35}~ gm/cm^3$ \\
        $\alpha$            &  $1.2138\times10^{-37}$                & $1.2687\times10^{-37}$ \\
        $q_{0}$             &  $-0.5276$                             & $-0.5756$ \\
        $z_{t}$             &  $0.6455$                              & $0.7356$ \\
        \hline
   \end{tabular}
  \caption{The present values of cosmological parameters along two data sets SNe Ia and Hubble Parameter $H(z)$.}\label{T3}
\end{table}
%%%%%%%%%%%%%%%%%%%%%%%%%%%%%%%%%%%%%%%%%%%%%%%%%%%%%%%%%%%%%%%%%%%%%%%%
\subsection*{Energy parameters}
The energy density parameters $\Omega_{(m)}$ and $\Omega_{(de)}$ is given by equation (\ref{eq13}) and its geometrical behaviour is shown in figure 4a \& figure 4b. One can see that as $z\to-1$, $(\Omega_{(m)}, \Omega_{(de)})\to(0, 1)$ and this reveals that the late-time universe is dark energy dominated and approaches to $\Lambda$CDM model, which is in good agreement with recent observations. In our model, the dark energy term is derived from perfect-fluid source and this shows the importance of the model. The present values of energy parameters are mentioned in Table 2 \& 3. The value of the model parameter $\alpha$ is estimated as $\alpha=1.26870954\times10^{-37}$ for $H(z)$ data and $\alpha=1.21383844\times10^{-37}$ for Pantheon data of SNe Ia which is compatible with recent values.
%%%%%%%%%%%%%%%%%%%%%%%%%%%%%%%%%%%%%%%%%%%%%%%%%%%%%%%%%%%%%%%%%%%%%%
\begin{figure}[H]
%\centering
	a.\includegraphics[width=8cm,height=7cm,angle=0]{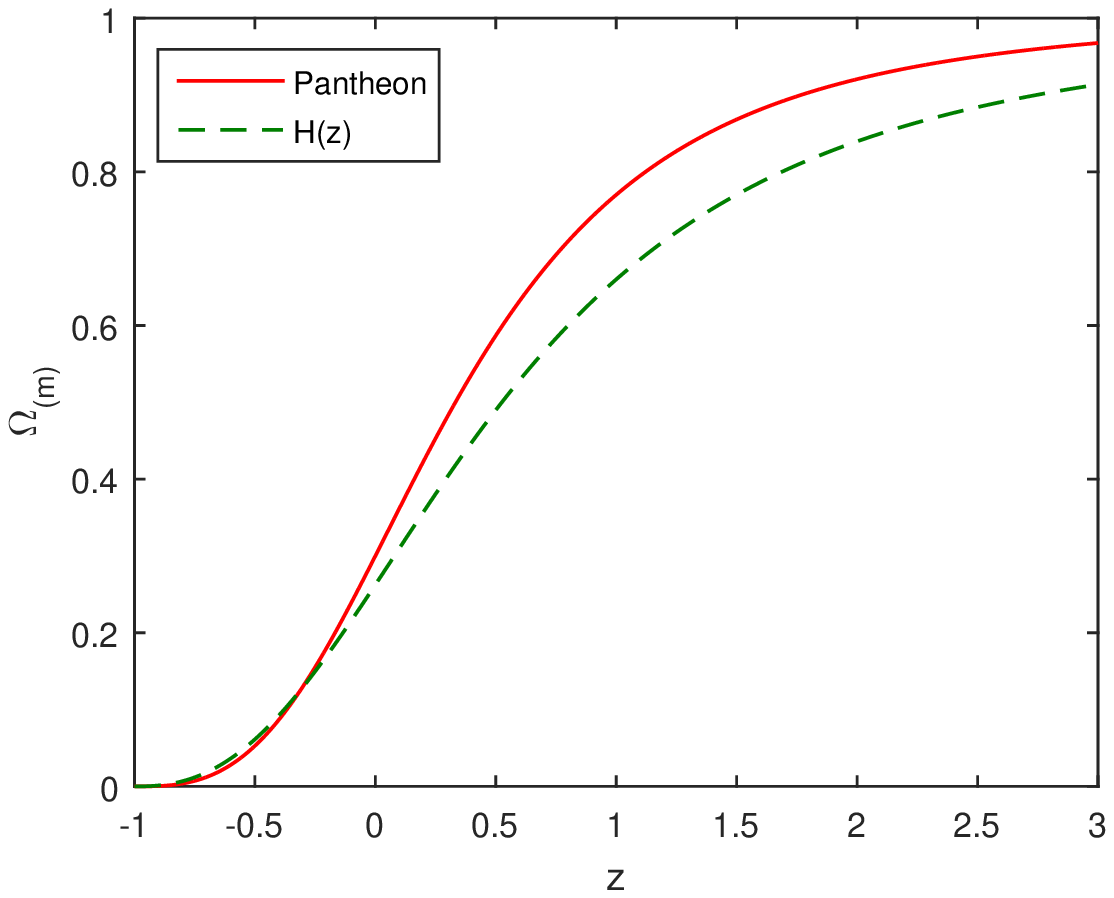}
    b.\includegraphics[width=8cm,height=7cm,angle=0]{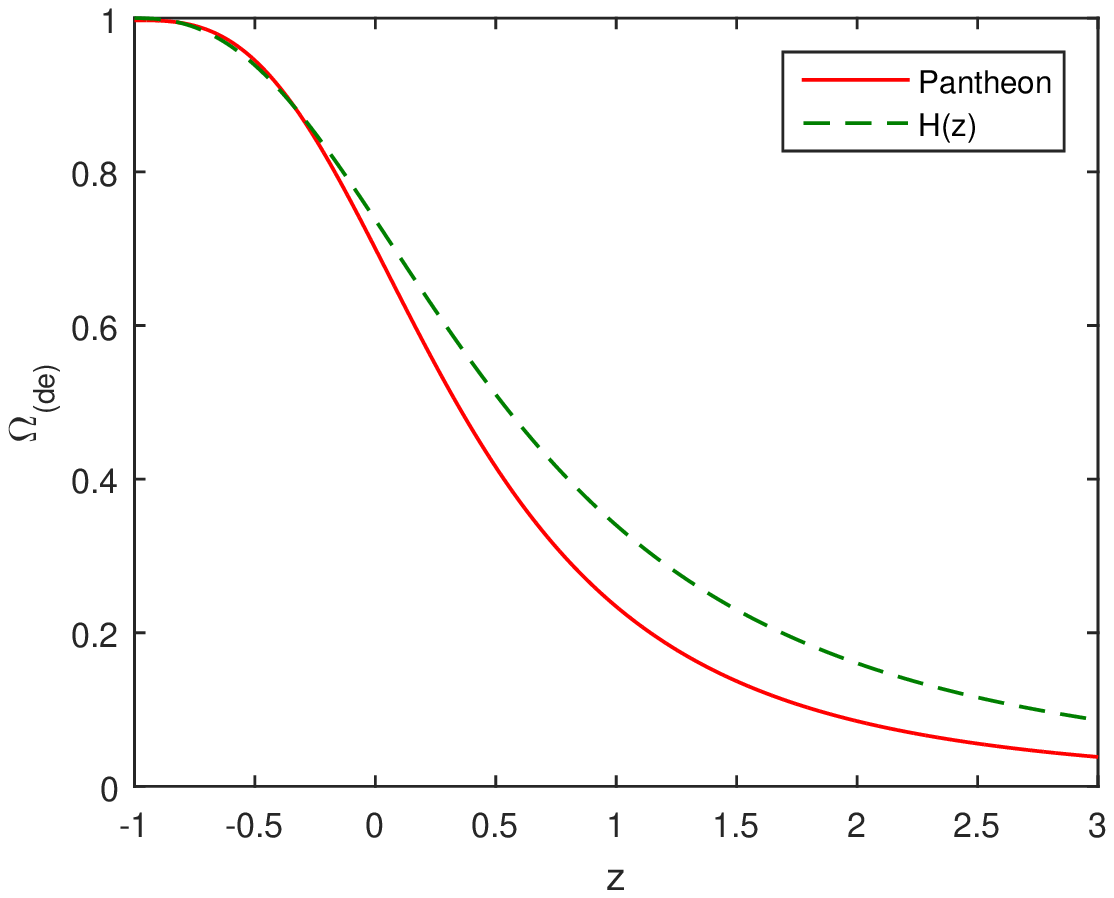}
  	\caption{The evolution of matter energy density parameter $\Omega_{(m)}$ and dark energy density parameter $\Omega_{(de)}$ over redshift $z$ respectively.}
\end{figure}
%%%%%%%%%%%%%%%%%%%%%%%%%%%%%%%%%%%%%%%%%%%%%%%%%%%%%%%%%%%%%%%%%%%%%%
The expression for dark-energy density and pressure is derived in equation (\ref{eq9}) and its geometrical behaviour is shown in figure 5a \& figure 5b. One can see that as $z\to-1$, the dark energy density $\rho^{(de)}$ increases and the negative pressure of dark energy $p^{(de)}$ is also increases. This shows that the present universe is dark energy dominated and this energy comes from matter fluid source which is responsible for acceleration in expansion.
%%%%%%%%%%%%%%%%%%%%%%%%%%%%%%%%%%%%%%%%%%%%%%%%%%%%%%%%%%%%%%%%%%%%%%
\begin{figure}[H]
%\centering
	a.\includegraphics[width=8cm,height=7cm,angle=0]{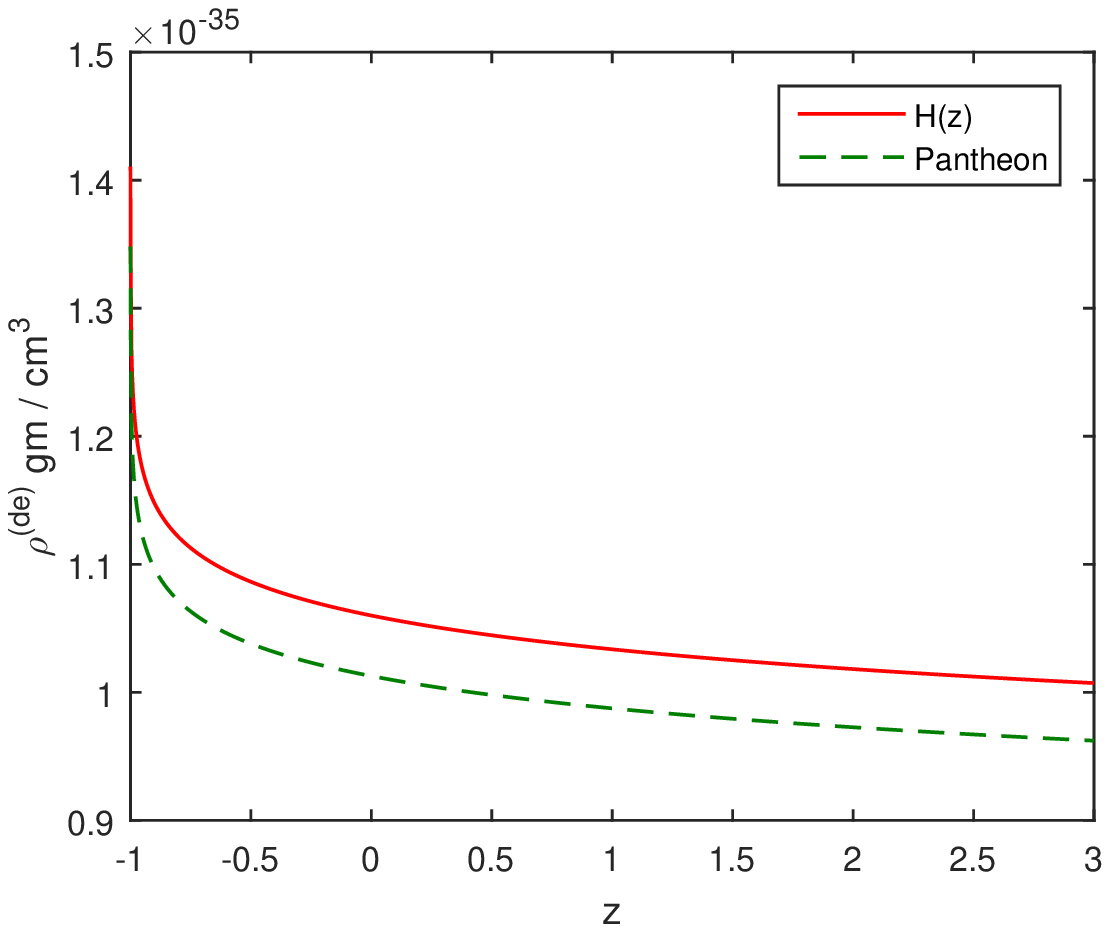}
    b.\includegraphics[width=8cm,height=7cm,angle=0]{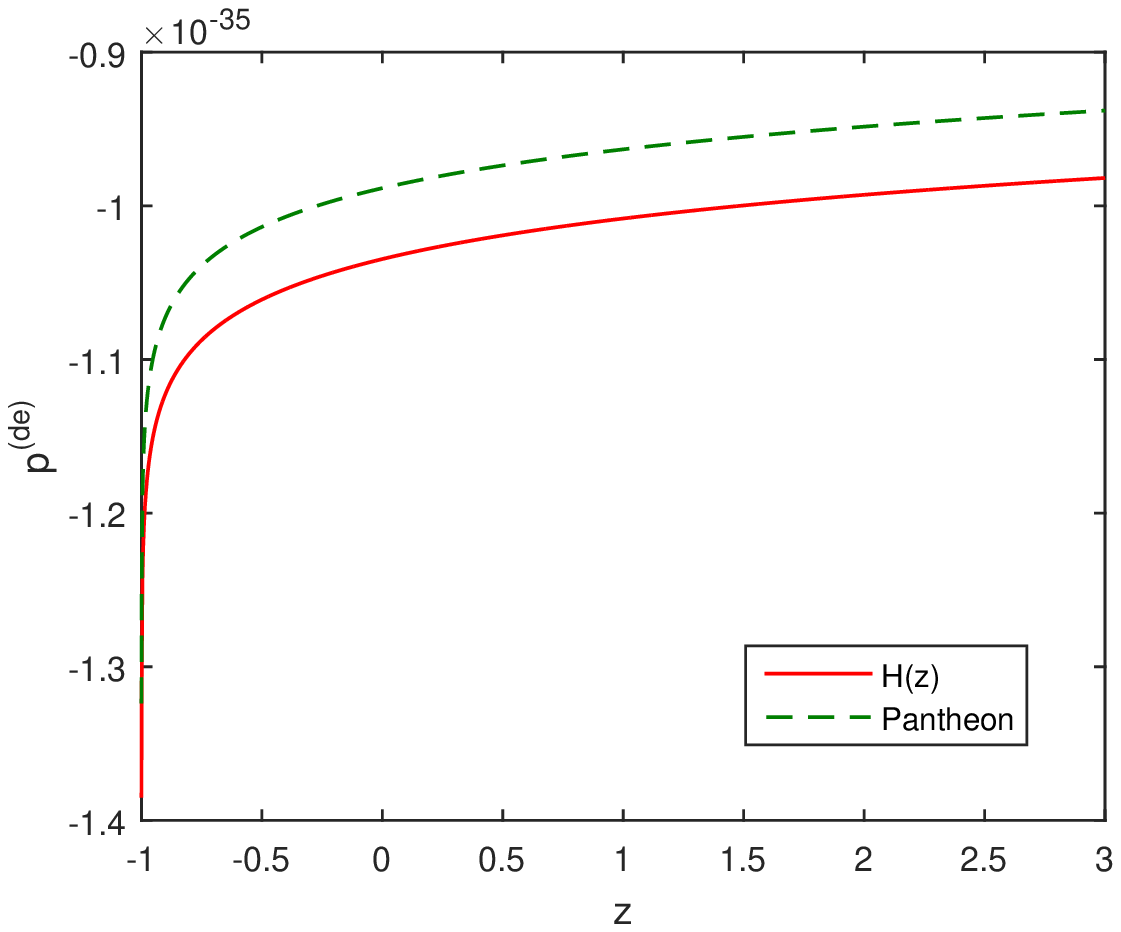}
  	\caption{The evolution of dark energy density $\rho^{(de)}$ and dark-fluid pressure $p^{(de)}$ over redshift $z$.}
\end{figure}
%%%%%%%%%%%%%%%%%%%%%%%%%%%%%%%%%%%%%%%%%%%%%%%%%%%%%%%%%%%%%%%%%%%%%%
\subsection*{Analysis of Om Diagnostic}
The cosmic dark energy models can be classified through behaviour of Om diagnostic function \cite{ref105}. The simplest diagnostic for a spatially flat universe is given by

\begin{equation}\label{eq24}
  Om(z)=\frac{\left(\frac{H(z)}{H_{0}}\right)^{2}-1}{(1+z)^{3}-1}
\end{equation}
where $H(z)$ is the Hubble parameter given in Eq.~(\ref{eq18}) and $H_{0}$ is its current value. A negative slope of $Om(z)$ corresponds to pith motion, and a positive slope corresponds to phantom motion. The $Om(z)$ constant represents the $\Lambda$CDM model.
%%%%%%%%%%%%%%%%%%%%%%%%%%%%%%%%%%%%%%%%%%%%%%%%%%%%%%%%%%%%%%%%%%%%%%
\begin{figure}[H]
\centering
	\includegraphics[width=9cm,height=7cm,angle=0]{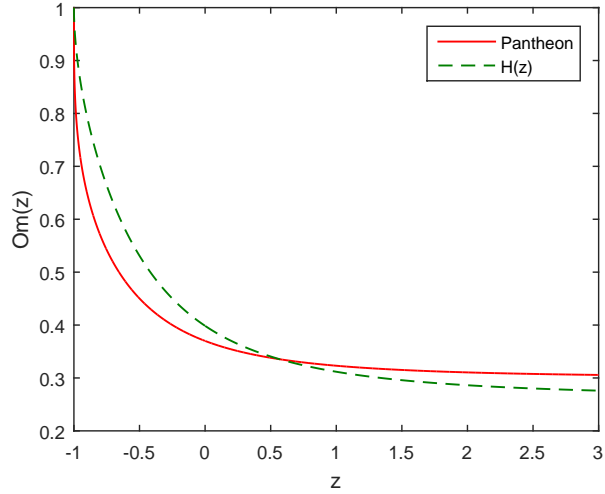}
  	\caption{The geometrical behaviour of $Om(z)$ function over redshift $z$.}
\end{figure}
%%%%%%%%%%%%%%%%%%%%%%%%%%%%%%%%%%%%%%%%%%%%%%%%%%%%%%%%%%%%%%%%%%%%%%
Figure 6 shows the geometrical behaviour of Om diagnostic function $Om(z)$ over redshift $z$ and mathematical expression is given in above equation (\ref{eq24}). From figure 6, one can see that the slope of $Om(z)$ function is negative for our model and it shows the quintessence behaviour of the model. Thus, the model derived in $f(R,T)=R-16\pi G \ln(T)$ gravity behaves just like quintessence dark energy model. Also, it is supported by the behaviour of dark energy EoS $\omega^{(de)}>-1$ as in our derived model $\omega^{(de)}=-0.81\pm0.22149, -0.58155\pm0.16941$ along two observational datasets Pantheon and $H(z)$ respectively. One can see that a quintessence dark energy model can be equivalently mapped to generalized holographic dark energy model with a suitable choice of cut-off, which is clearly shown in \cite{ref106}. The behavior of our derived model is quintessential and hence, it can be equivalently mapped to generalized holographic dark energy model with a suitable choice of cut-off and this shows the viability of the model.\\

%%%%%%%%%%%%%%%%%%%%%%%%%%%%%%%%%%%%%%%%%%%%%%%%%%%%%%%%%%%%%%%%%%%%%%%%%%%%%%%%%%%%%%%%%%%%%%%%%%
%%%%%%%%%%%%%%%%%%%%%%%%%%%%%%%%%%%%%%%%%%%%%%%%%%%%%%%%%%%%%%%%%%%%%%%%%%%%%SECTION 5
%%%%%%%%%%%%%%%%%%%%%%%%%%%%%%%%%%%%%%%%%%%%%%%%%%%%%%%%%%%%%%%%%%%%%%%%%%%%%%%%%%%%%%%%%%%%%%%%%%%%%
\section{Conclusion}
The present research paper is an investigation of dark energy nature of logarithmic $f(R, T)$-gravity cosmology in a flat FLRW space-time universe. We have derived modified Einstein's field equations for the function $f(R, T)=R-16\pi G\alpha\ln(T)$ where $R$ is the Ricci scalar curvature, $T$ is the trace of the stress energy momentum tensor and $\alpha$ is a model parameter. We have solved field equations in the form of two fluid scenario as perfect-fluid and dark-fluid, where dark fluid term is derived in the form of perfect fluid source. We have made an observational constraints on the cosmological parameters $\Omega_{(m)}, \omega^{(de)}$ and $H_{0}$ using $\chi^{2}$ test with observational datasets like Pantheon sample of SNe Ia and $H(z)$. With these constraints we have discussed our model with deceleration parameter $q$, energy parameters $\Omega_{(m)}, \Omega_{(de)}$, EoS parameter $\omega^{(de)}$ etc. and Om diagnostic function. The main features of the derived model are as follows:
\begin{itemize}
  \item The derived model shows a transit phase (decelerating to accelerating) model with present values $q_{0}=-0.5276, -0.5756$ along two observational datasets Pantheon and $H(z)$ respectively.
  \item The transition redshift is estimated as $z_{t}=0.6455, 0.7356$ for two data sets Pantheon and $H(z)$ respectively, which is in good agreement with recent observations \cite{ref102}.
  \item The present values of energy parameters are estimated as $\Omega_{(m)}=0.2981\pm0.08921$, $\Omega_{(de)}=0.7019$ for Pantheon data and $\Omega_{(m)}=0.26535\pm0.01254$, $\Omega_{(de)}=0.73465$ for $H(z)$ datasets.
  \item The behaviour of dark energy EoS $\omega^{(de)}>-1$ as in our derived model $\omega^{(de)}=-0.81\pm0.22149, -0.58155\pm0.16941$ along two observational datasets respectively.
  \item The values of model parameter $\alpha$ are estimated as $\alpha=1.26870954\times10^{-37}$ for $H(z)$ data and $\alpha=1.21383844\times10^{-37}$ for Pantheon data of SNe Ia which is compatible with recent values.
  \item The derived $f(R, T)$ model shows a quintessence dark energy model $\omega^{(de)}>-1$ and late-time universe approaches to $\Lambda$CDM model.
  \item We have estimated the present value of Hubble constant as $H_{0}=67.67851\pm0.86949$ Km/s/Mpc that may resolve Hubble tension issues in cosmology, since it is consistent with both early and late-time universe.
\end{itemize}
Thus, the derived cosmological model behaves as an quintessence dark energy model and the dark energy term is derived from perfect fluid source which is an interesting feature of this model.
%%%%%%%%%%%%%%%%%%%%%%%%%%%%%%%%%%%%%%%%%%%%%%%%%%%%%%%%%%%%%%%%%%%%%%%%%%%%%%%%%%%%%%%%%%%%%
%%%%%%%%%%%%%%%%%%%%%%%%%%%%%%%%%%%%%%%%%%%%%%%%%%%%%%%%%%%%%%%%%%%%
%%%%%%%%%%%%%%%%%%%%%%%%%%%%%%%%%%%%%%%%%%%%%%%%%%%%%%%%%%%%%%%%%%%%%%%%%%%%%%%%%%%%%%%%%%%%%%
\section*{Acknowledgement}
We are thankful to reviewers and editors for their motivational suggestions to improve our manuscript.
%%%%%%%%%%%%%%%%%%%%%%%%%%%%%%%%%%%%%%%%%%%%%%%%%%%%%%%%%%%%%%%%%%%%%%%%%%%%%%%%%%%%%%%%%%%%%%%%%%%%
%%%%%%%%%%%%%%%%%%%%%%%%%%%%%%%%%%%%%%%%%%%%%%%%%%%%%%%%%%%
%%%%%%%%%%%%%%%%%%%%%%%%%%%%%%%%%%%%%%%%%%%%%%%%%%%%%%%%%%%%%%%%%%%%%%%%%%%%%%%%%%%%%%%%%%%%%%%%%%%%%%
\section{Declarations}
\subsection*{Funding and/or Conflicts of interests/Competing interests}
The authors of this article have no conflict of interests. Also, this work is not supported by any type of funding sources.
%%%%%%%%%%%%%%%%%%%%%%%%%%%%%%%%%%%%%%%%%%%%%%%%%%%%%%%%%%%%%%%%%%%%%%%%%%%%%%%%%%%%%%%%%%%%%%%%%%%%
%%%%%%%%%%%%%%%%%%%%%%%%%%%%%%%%%%%%%%%%%%%%%%%%%%%%%%%%%%%
%%%%%%%%%%%%%%%%%%%%%%%%%%%%%%%%%%%%%%%%%%%%%%%%%%%%%%%%%%%%%%%%%%%%%%%%%%%%%%%%%%%%%%%%%%%%%%%%%%%%%%
\section{Data Availability}
We have not used any data for the analysis presented in this work.
%%%%%%%%%%%%%%%%%%%%%%%%%%%%%%%%%%%%%%%%%%%%%%%%%%%%%%%%%%%%
	
\end{document}